\documentclass[doublecol]{epl2}
\usepackage{amsmath}
\usepackage{amssymb}
\usepackage{overpic}
\usepackage[table]{xcolor}

\newcommand{\tauK}{\tau_\eta}
\newcommand{\etaK}{\eta}

\newcommand{\zzhat}{\hat{\ve z}}
\newcommand{\vs}{v_{{\rm s}}}
\newcommand{\tausurf}{\tau^{{\rm surf}}}

\newcommand{\ve}[1]{\ensuremath{\mbox{\boldmath$#1$}}}
\newcommand{\ma}[1]{\ensuremath{\mathbb{#1}}}

\title{Guided navigation in turbulence}
\shorttitle{}

\author{J. Qiu\inst{1} \and K. Gustavsson\inst{1}}
\shortauthor{J. Qiu \etal}

\institute{
  \inst{1} Department of Physics, Gothenburg University, Gothenburg, SE-40530 Sweden
}

\abstract{
Navigation in turbulence is challenging because local flow measurements provide limited information about favorable paths beyond the flow correlation scales.  We investigate vertical navigation by swimmers in homogeneous isotropic turbulence and compare strategies using local flow information with strategies guided by high-performing trajectories. Compared with naive upward swimming, local strategies enhance the mean vertical velocity by about $10\%$ of the root-mean-square flow velocity, while guidance by precalculated trajectories reaches enhancements of about $40\%$. This gain is limited by the time required to intercept favorable trajectories, particularly for slow swimmers. We then show that precalculated trajectories are unnecessary: swimmers can dynamically identify high-performing members of a swarm and use them as targets. For sufficiently fast swimmers, this collective strategy keeps much of the benefit of precalculated guidance using only the instantaneous swarm state. Our results demonstrate how distributed trajectory information can be used to overcome limitations of local sensing for navigation in turbulence.
}

\begin{document}

\maketitle

\section{Introduction}
Navigation in complex flows is important in biology and engineering. Examples include zooplankton performing long-distance vertical migrations~\cite{hays2003review}, fish adapting their motion to turbulent vortex streets~\cite{liao2003karman}, and autonomous aerial and ocean vehicles using flow measurements for path planning~\cite{chan2023wind,vagale2021path}.

Most approaches for navigation in complex flows rely on spatially local flow information. Short-time expansions use local flow signals to derive navigation strategies~\cite{monthiller2022surfing,mousavi2024efficient,qiu2026adaptive}, but may be suboptimal over longer horizons~\cite{mousavi2025short}.
Reinforcement-learning approaches optimize over longer horizons~\cite{Sutton1998reinforcement,Mehlig2021}, but most existing studies likewise rely on local flow observations~\cite{colabrese2017flow,verma2018Efficient,Alageshan2020machine,gunnarson2021learning,qiu2022navigation,qiu2022active,godavarthi2025leveraging,jiao2025sensing,koh2025physicsguided,xu2026learning}.
Temporal memory of local observations can further improve navigation, for example through reinforcement learning or Bayesian inference~\cite{rando2025qlearning,heinonen2023optimal}, and strategies under partial observability can exploit intermittent turbulent cues~\cite{rigolli2022alternation}. Nevertheless, such information provides limited knowledge of favorable paths in distant regions of the flow. This limitation is particularly relevant in turbulence, which spans a broad range of spatial and temporal scales.

Nonlocal information could enable longer-range path planning. Optimal-control approaches, dating back to Zermelo's navigation problem~\cite{Zermelo1931uber}, use global flow information to identify efficient trajectories~\cite{Liebchen2019optimal,Schneider2019optimal,calascibetta2023optimal ,piro2024} and to track moving targets in turbulence~\cite{calascibetta2023optimal}.
But such information is generally unavailable to autonomous agents.
Even when available, finding optimal trajectories in turbulence is challenging because sensitivity to initial conditions makes long-time trajectory optimization increasingly difficult~\cite{Biferale2019}. This motivates the question of whether favorable trajectories can be identified and exploited without globally optimizing the swimmer path.

A swarm can provide nonlocal information since different agents simultaneously sample trajectories in different regions of the flow.
In turbulent olfactory search, sharing information about the motion of other agents can substantially improve collective search~\cite{durve2020collective}. More generally, information sharing can improve collective search, rendezvous, and navigation~\cite{mcguire2019minimal,yang2022autonomous,li2026multi}, while distinct agent roles can provide an advantage by balancing exploration and exploitation~\cite{piro2026policy}.
These approaches primarily concern finding a prescribed target, while our problem concerns exploiting favorable, dynamically evolving paths through the flow.
How distributed trajectory information can be used for navigation in multiscale turbulence remains an open question.

Here, we address this problem by using high-performing swimmers as moving targets for other swimmers. We first establish the potential of nonlocal trajectory information using precalculated trajectories and then develop a swarm strategy in which targets are selected dynamically. We show that target guidance substantially outperforms local strategies when favorable trajectories can be intercepted sufficiently rapidly. For fast swimmers, dynamically selected targets retain much of this advantage without precalculated trajectories, demonstrating how a swarm can exploit nonlocal information for turbulent navigation.

\section{Dynamics}

Each swimmer evolves according to
\begin{align}
\dot{\ve x}_t&=\ve u(\ve x_t,t)+\vs\ve n_t^{\rm ctrl}\,.
\label{eq:eom}
\end{align}
Here, $\ve x$ is the swimmer position, $\ve u$ the flow velocity, and $\vs\ve n_t^{\rm ctrl}$ its swimming velocity, with $\ve n_t^{\rm ctrl}$ a unit vector.
The swimmer controls $\ve n_t^{\rm ctrl}$ instantaneously, aiming to navigate along the vertical direction $\zzhat$ based on available information.
Interactions between swimmers are neglected.

The flow $\ve u$ is taken from stationary, forced homogeneous isotropic turbulence in the JHTDB \texttt{isotropic1024coarse} dataset~\cite{Li2008,Minping2012}. The data were obtained by direct numerical simulation of the incompressible Navier-Stokes equations in a triply periodic domain on a $1024^3$ grid using a pseudo-spectral method. 
The Taylor-scale Reynolds number is $Re_\lambda\approx433$, the periodic box size is $L_{\rm box}=2200\etaK$ and the simulation spans $T\approx237\tauK$, or about five large-eddy turnover times, where $\etaK$ and $\tauK$ are the Kolmogorov length and time.

To integrate Eq.~(\ref{eq:eom}), we sample the flow velocity and gradients at intervals $\Delta t\approx0.05\tauK$ and use Adams--Bashforth integration with time step $0.1\Delta t$, linearly interpolating the flow between samples.

\section{Baseline local strategies}
The dynamics (\ref{eq:eom}) combines flow advection and swimming. The swimming contribution to vertical motion is maximized by heading upwards,
\begin{align}
\ve n^{\rm up}&=\zzhat\,.
\label{eq:ctrl_up}
\end{align}
By contrast, the local flow contribution $u_z$ is maximized by swimming along its gradient, $\ve n^{\rm GA}=\ve\nabla u_z/|\ve\nabla u_z|$, parameterized by polar and azimuthal angles $\theta^{\rm GA}$ and $\varphi^{\rm GA}$. 
To balance these objectives, we use clamped gradient ascent,
\begin{align}
\varphi^{\rm CGA}=\varphi^{\rm GA}
\hspace{0.1cm}\mbox{and}\hspace{0.1cm}
\theta^{\rm CGA}=\left\{
\begin{array}{ll}
\!\!\theta^{\rm GA} & \!\!\mbox{if }\theta^{\rm GA}<\theta_{\rm c}\cr
\!\!\theta_{\rm c} & \!\!\mbox{otherwise}
\end{array}
\right.\,,
\label{eq:ctrl_cga}
\end{align}
which follows gradient ascent while enforcing a vertical swimming component of at least $\cos\theta_{\rm c}$. We optimize $\theta_{\rm c}$ numerically for each $\vs$.

Based on a short-time local flow expansion, Monthiller et al.~\cite{monthiller2022surfing} proposed the surfing strategy
\begin{align}
\ve n^{\rm surf}&=\frac{\exp[\ma A^{\sf T}\tausurf]\zzhat}{|\exp[\ma A^{\sf T}\tausurf]\zzhat|}\,.
\label{eq:ctrl_surf}
\end{align}
where $\ma A$ is the flow-gradient matrix and the free parameter $\tausurf$ is optimized numerically. 
For small $\tausurf$,
\begin{align}
\ve n^{\rm surf}\sim \frac{\zzhat+\tausurf\ve\nabla u_z}{|\zzhat+\tausurf\ve\nabla u_z|}\,.
\end{align}
showing that $\tausurf$ controls the trade-off between upward swimming and gradient ascent. For larger $\tausurf$, this interpretation no longer applies.

We use upward swimming (\ref{eq:ctrl_up}), clamped gradient ascent (\ref{eq:ctrl_cga}), and surfing (\ref{eq:ctrl_surf}) as baselines. Performance is quantified by the dimensionless vertical velocity enhancement
\begin{align}
\chi_z=\frac{1}{T}\int_0^T{\rm d}t'\frac{v_z(t')-\vs}{u_{\rm rms}}\,,
\label{eq:chi}
\end{align}
where $u_{\rm rms}=\langle\ve u^2\rangle^{1/2}$ and $T$ is the simulation time. 
Thus, if $\chi_z>0$, the swimmer exploits the flow to achieve a mean vertical velocity exceeding its swimming speed.

\begin{figure}
\includegraphics[width=\linewidth]{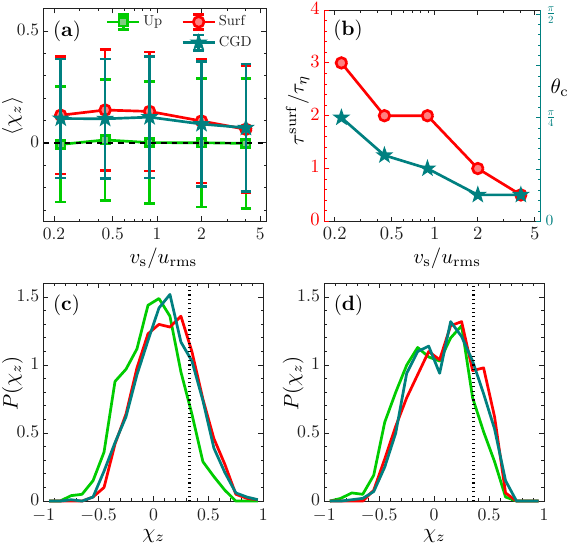}
\caption{
\label{fig:vzMeanLocal}
({\bf a}) Average of vertical velocity enhancement $\langle\chi_z\rangle$ (\ref{eq:chi}) against swimming speed $\vs$ for $\ve n^{\rm up}$ (\ref{eq:ctrl_up}) [{\tiny$\blacksquare$},green], $\ve n^{\rm CGA}$ (\ref{eq:ctrl_cga}) [$\star$,teal], and $\ve n^{\rm surf}$ (\ref{eq:ctrl_surf}) [$\bullet$,red].
Points show averages over 1000 trajectories with random initial positions and bars $\pm$ one standard deviation.
The dashed line shows free vertical swimming without flow, $\chi_z=0$.
({\bf b}) Optimal $\theta_{\rm c}$ and $\tausurf$ for $\ve n^{\rm CGA}$ and $\ve n^{\rm surf}$, respectively.
({\bf c},{\bf d}) Distributions of $\chi_z$ for the baseline strategies with ({\bf c}) $\vs/u_{\rm rms}=0.22$ and ({\bf d}) $2$.
Dotted lines show thresholds for selecting high-performing swimmers.
}
\end{figure}

\subsection{Performance of baseline strategies}
Figure~\ref{fig:vzMeanLocal}({\bf a}) shows the mean and standard deviation of $\chi_z$ for the three baseline strategies.
Clamped gradient ascent and surfing use the numerically optimized parameters $\theta_{\rm c}$ and $\tausurf$ shown in Fig.~\ref{fig:vzMeanLocal}({\bf b}).
Both decrease with increasing $\vs$, reflecting the growing importance of swimming relative to advection.

Upward swimming, $\ve n^{\rm up}$, gives $\langle\chi_z\rangle\approx0$ and fails to exploit the flow.
In contrast, clamped gradient ascent $\ve n^{\rm CGA}$ and surfing $\ve n^{\rm surf}$ preferentially sample upwelling regions and achieve comparable enhancements. 
Surfing performs better at small and moderate $\vs$, while clamped gradient ascent is slightly better at the largest $\vs$. 
Our surfing results agree with Ref.~\cite{monthiller2022surfing} and extend to larger $\vs$.

The standard deviations of $\chi_z$ [bars in Fig.~\ref{fig:vzMeanLocal}({\bf a})] are of order $0.3$, much larger than the means, indicating strong trajectory-to-trajectory variability.
For $\vs\ll u_{\rm rms}$, many swimmers even have negative mean vertical velocities despite a positive ensemble average.
By contrast, the uncertainty in $\langle\chi_z\rangle$ is of order $0.01$, below the marker size.

Figures~\ref{fig:vzMeanLocal}({\bf c},{\bf d}) show $P(\chi_z)$ for $\vs/u_{\rm rms}=0.22$ and $2$.
For all strategies, including upward swimming despite its near-zero mean, some initial positions perform well above the mean by preferentially sampling positive vertical flow, producing enhancements up to order unity.

Chaotic dynamics may bring poorly performing swimmers close to high-performing trajectories. 
By adjusting its swimming direction, a swimmer could intercept and track such trajectories, substantially improving its performance. This motivates strategies that target high-performing trajectories, potentially shifting $P(\chi_z)$ toward larger values while narrowing its distribution.

\section{Target-guided strategies}
\label{sec:target_tracking}
We construct a population of target trajectories for each $\vs$ by selecting those with $\langle v_z\rangle$ above a threshold from all trajectories generated by Eqs.~(\ref{eq:ctrl_up}), (\ref{eq:ctrl_cga}), and (\ref{eq:ctrl_surf}), including all tested $\theta_{\rm c}$ and $\tausurf$.
The thresholds, indicated by dotted lines in Fig.~\ref{fig:vzMeanLocal}({\bf c},{\bf d}), are chosen to yield $N^{\rm target}\sim1000$ targets per $\vs$. Combining strategies and parameters increases target diversity and reduces correlations among trajectories.

We consider two target-guided controls: greedy pursuit of the nearest target and predictive interception of the most promising target.
At each time step $\Delta t$, the control evaluates the minimum-image separation $\ve R_i$ from the swimmer at $\ve x$ to each target at $\ve x_i$.
If $R_{i,z}>100\etaK$, the target is deemed unreachable, and the image one box length $L_{\rm box}$ below is used instead.

Under greedy control, the swimmer dynamically aligns toward the nearest target:
\begin{align}
\ve n^{\rm greedy}=\frac{\ve R_{i^\star}}{|\ve R_{i^\star}|}\,,\hspace{0.5cm}i^\star = \arg\min_{i} |\ve R_{i}|\,.
\label{eq:ctrl_greedy}
\end{align}

\begin{figure}
\centering
\includegraphics[width=0.35\textwidth]{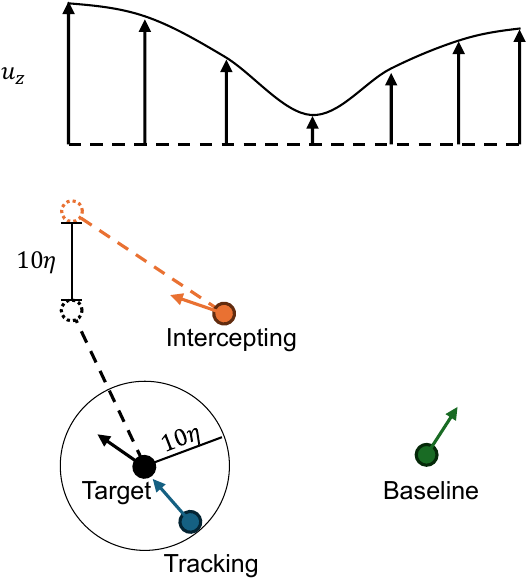}
\caption{
\label{fig:schematic}
Predictive-control modes. Black: target swimmer. Blue: tracking a nearby target ($|\ve R|<10\etaK$). Orange: intercepting above the predicted target position. Green: baseline control when interception offers no predicted improvement.
}
\end{figure}

Predictive control has three modes (Fig.~\ref{fig:schematic}). 
The closest target within $10\etaK$ is tracked by swimming toward it, $\ve n^{\rm track}=\ve n^{\rm greedy}$. The distance $10\etaK$ estimates the dissipative range where the flow is spatially smooth~\cite{pope2000turbulent,bec2024statistical}. 
If no target lies within $10\etaK$, a target is selected for interception. To predict future motion while suppressing fluctuations, the flow and target velocities are estimated from their past values using exponential smoothing with time scale $3.5\tauK$.
Assuming these velocities remain constant during interception, the interception direction $\ve n_i^{\rm int}$ and time $\tau_i$ are obtained analytically for each of the 50 nearest targets. 
This lead-interception approach anticipates the target position rather than pursuing its instantaneous position~\cite{fajen2004visual,isaacs1965differential,zheng2021timeoptimal}. The swimmer aims at the highest reachable point less than $10\etaK$ above the target, providing a margin against flow fluctuations.

Each reachable target is ranked by its predicted vertical displacement until the end of the simulation,
\begin{align}
J_i=\overline{u}_z\min(\tau_i,T_{\rm E})
+\vs n^{\rm int}_{i,z}\tau_i
+z_i(T)-z_i(t+\tau_i)\,,
\label{eq:Ji}
\end{align}
where $\overline{u}_z$ is the smoothed vertical flow velocity and $T_{\rm E}$ the Eulerian integral time. 
The first two terms estimate displacement before interception and the last gives the target displacement thereafter. 
The interception direction for the best target is
\begin{align}
\ve n^{\rm int}=\ve n^{\rm int}_{i^\star}\,,\hspace{0.5cm}i^\star=\arg\max_i J_i\,.
\label{eq:ctrl_int}
\end{align}
Interception is initiated only if $J_{i^\star}>J_{\rm stay}$, where
\begin{equation}
J_{\rm stay}=\overline{u}_z\min(T_{\rm E},T-t)+\vs(T-t)
\label{eq:Jstay}
\end{equation}
estimates the displacement from staying in the current flow and swimming upward.
Otherwise, the swimmer uses a baseline strategy, $\ve n^{\rm baseline}$, chosen as the better of surfing and clamped gradient ascent at the given $\vs$.
In summary,
\begin{align}
\ve n^{\rm pred}=
\begin{cases}
\ve n^{\rm track}
& \mbox{if } \min_i|\ve R_i|<10\etaK\,,\cr
\ve n^{\rm int}
& \mbox{if } \max_i J_i>J_{\rm stay}\,,\cr
\ve n^{\rm baseline}
& \mbox{otherwise}\,.
\end{cases}
\label{eq:ctrl_pred}
\end{align}

\subsection{Performance of guided strategies}

\begin{figure}
\includegraphics[width=\linewidth]{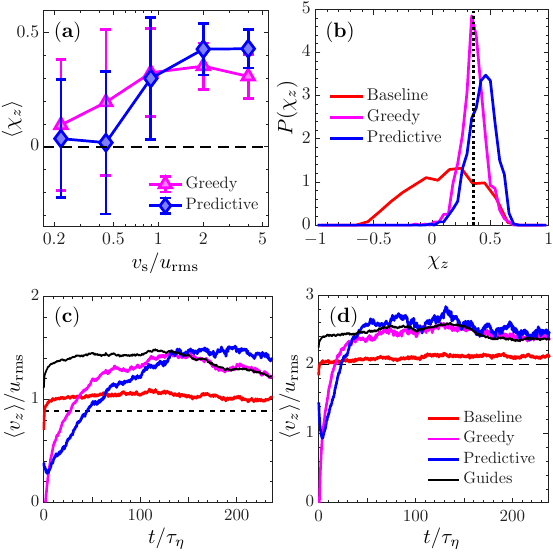}
\caption{
\label{fig:vzMeanGlobal}
({\bf a},{\bf b}) As in Fig.~\ref{fig:vzMeanLocal}({\bf a},{\bf d}) for guided strategies: $\ve n^{\rm greedy}$ [Eq.~(\ref{eq:ctrl_greedy});{\footnotesize$\blacktriangle$},magenta] and $\ve n^{\rm pred}$ [Eq.~(\ref{eq:ctrl_pred});{\scriptsize$\blacklozenge$},blue].
({\bf c},{\bf d}) Evolution of $\langle v_z(t)\rangle$, averaged over 1000 trajectories, for ({\bf c}) $\vs=0.89u_{\rm rms}$ and ({\bf d}) $2u_{\rm rms}$, using $\ve n^{\rm greedy}$ (magenta), $\ve n^{\rm pred}$ (blue), and $\ve n^{\rm baseline}=\ve n^{\rm surf}$ [Eq.~(\ref{eq:ctrl_surf});red]. 
Solid and dashed black lines show the average over target trajectories and $\vs$, respectively.
}
\end{figure}

Figure~\ref{fig:vzMeanGlobal}({\bf a}) shows the mean vertical velocity enhancement for the two guided strategies.
Predictive control outperforms the baselines for $\vs\gtrsim u_{\rm rms}$, reaching enhancements up to five times that of the best baseline in Fig.~\ref{fig:vzMeanLocal}({\bf a}), but performs worse at smaller $\vs$.
Greedy control gives smaller enhancements for $\vs>u_{\rm rms}$ but performs better at smaller $\vs$, outperforming the best baseline down to $\vs\sim 0.3u_{\rm rms}$.

For $\vs\ge u_{\rm rms}$, the standard deviations of $\chi_z$ are smaller than the mean enhancements.
Accordingly, the distributions for $\vs=2u_{\rm rms}$ [Fig.~\ref{fig:vzMeanGlobal}({\bf b})] are substantially narrower than for the baseline distribution.
Thus, the enhancement occurs for most initial conditions rather than being dominated by exceptionally successful trajectories.

Figures~\ref{fig:vzMeanGlobal}({\bf c},{\bf d}) show the evolution of $\langle v_z\rangle$ for $v_s/u_{\rm rms}=0.89$ and $2$.
Initially, guided swimmers select distant targets and underperform the baseline, but $\langle v_z\rangle$ increases as targets are intercepted. 
This transition is rapid for large $\vs/u_{\rm rms}$ but slows as $\vs$ decreases, particularly for predictive control.
At $\vs/u_{\rm rms}=0.44$, predictive control reaches high $\langle v_z\rangle$ only near the simulation end (not shown), leaving insufficient time to accumulate displacement.

For large times, $\langle v_z\rangle$ of greedy control approaches that of the target population (black), consistent with selecting targets by proximity rather than performance.
Predictive control exceeds $\langle v_z\rangle$ of the targets by preferentially selecting targets with high predicted returns. 
Both eventually outperform the baseline if interception is sufficiently rapid.
Thus, guided performance depends not only on target quality, but also on interception time.

\subsection{Interception and tracking dynamics}

\begin{figure*}
\includegraphics[width=\linewidth]{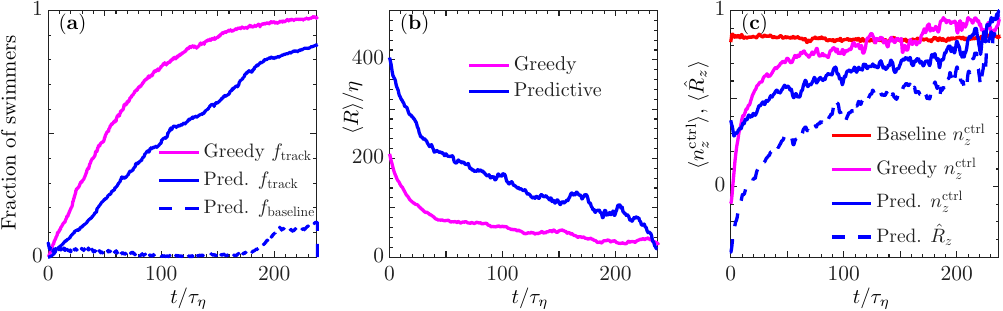}
\caption{
\label{fig:statisticsGlobal}
Interception and tracking statistics for 1000 swimmers with $\vs=0.89u_{\rm rms}$ using $\ve n^{\rm greedy}$ [Eq.~(\ref{eq:ctrl_greedy}); magenta], $\ve n^{\rm pred}$ [Eq.~(\ref{eq:ctrl_pred}); blue] and $\ve n^{\rm baseline}$ [Eq.~(\ref{eq:ctrl_surf}); red].
({\bf a}) Fractions in tracking, $f_{\rm track}$ (solid), and baseline, $f_{\rm baseline}$ (dashed), modes.
({\bf b},{\bf c}) Statistics for swimmers in interception mode, $f_{\rm int}=1-f_{\rm track}-f_{\rm baseline}$:
({\bf b}) mean target distance $\langle R\rangle$;
({\bf c}) mean vertical components of control, $\langle n^{\rm ctrl}_z\rangle$ (solid), and target direction, $\langle\hat R_z\rangle$ (dashed).
Data in ({\bf b},{\bf c}) are smoothed over~$\tauK$.
}
\end{figure*}

Figure~\ref{fig:statisticsGlobal} shows interception and tracking statistics for $\vs=0.89u_{\rm rms}$. 
A greedy swimmer tracks when its target is within $10\etaK$ and intercepts otherwise. Figure~\ref{fig:statisticsGlobal}({\bf a}) shows the fractions in tracking and baseline modes, $f_{\rm track}$ and $f_{\rm baseline}$, with $f_{\rm baseline}=0$ for greedy control. 
The remaining fraction, $f_{\rm int}=1-f_{\rm track}-f_{\rm baseline}$, is in interception mode. Greedy swimmers reach tracking faster than predictive swimmers, consistent with the faster increase of $\langle v_z\rangle$ in Fig.~\ref{fig:vzMeanGlobal}({\bf c}). 
Predictive swimmers use the baseline only near the simulation end, when interceptions become unfavorable.

Most swimmers eventually track for $\vs=0.89u_{\rm rms}$, with $f_{\rm track}\sim1$.
At smaller $\vs$, the final tracking fraction decreases, while $f_{\rm baseline}$ increases for predictive control (not shown). These trends confirm that slow target interception causes the reduced guided-strategy performance at small $\vs$ in Fig.~\ref{fig:vzMeanGlobal}({\bf a}).

Figures~\ref{fig:statisticsGlobal}({\bf b},{\bf c}) characterize swimmers during interception. Predictive swimmers select substantially more distant targets than greedy swimmers [Fig.~\ref{fig:statisticsGlobal}({\bf b})], initially by about a factor of two. 
This follows from their selection rules: greedy swimmers choose the nearest target, whereas predictive swimmers may accept more distant targets for higher predicted returns. This selectivity also explains why predictive swimmers initially have positive $\langle v_z\rangle$, while greedy swimmers have $\langle v_z\rangle\approx0$ [Fig.~\ref{fig:vzMeanGlobal}({\bf c})].
The mean target distance decreases as swimmers approach their targets. Once tracking begins, they no longer contribute to the interception-mode average $\langle R\rangle=\langle|\ve R|\rangle$ in Fig.~\ref{fig:statisticsGlobal}({\bf b}).
Near the simulation end, the sharp decrease for predictive control reflects swimmers reverting to baseline when no target has sufficient predicted return.

Figure~\ref{fig:statisticsGlobal}({\bf c}) shows the mean vertical control component, $\langle n_z^{\rm ctrl}\rangle$, and swimmer-to-target direction, $\langle\hat R_z\rangle=\langle R_z/R\rangle$, with the latter omitted for greedy control because $\ve n^{\rm ctrl}=\hat{\ve R}$. 
Initially, greedy swimmers pursue nearly isotropically distributed targets, giving $\langle n_z^{\rm ctrl}\rangle\approx0$. It later becomes positive as the remaining intercepting swimmers mainly approach targets from below.

Predictive control behaves differently: it initially selects targets below the swimmer, $\langle\hat R_z\rangle<0$, while maintaining upward-biased swimming, $\langle n_z^{\rm ctrl}\rangle>0$. More generally, $\langle n_z^{\rm ctrl}\rangle>\langle\hat R_z\rangle$ throughout. Predictive swimmers therefore balance interception with upward swimming rather than simply pointing toward their targets.

\section{Swarm navigation}
Target-guided strategies show that nonlocal information can outperform optimized local control.
In practice, however, precalculating target trajectories may be infeasible.
We therefore consider swarm navigation, dynamically dividing a population of swimmers into targets and pursuers. 
At each time step $\Delta t$, a fraction $p_{\rm target}$ with the highest smoothed $u_z$ is selected as targets and follows the baseline strategy. 
The remaining swimmers pursue these targets using a modified predictive control.

Pursuers predict over a horizon $T_{\rm p}\le T_{\rm E}$, where $T_{\rm E}$ is the integral time scale. The predicted total vertical displacement associated with selecting target $i$ in Eq.~(\ref{eq:Ji}) is replaced by
\begin{align}
J_i=(\overline{u}_z+\vs n^{\rm int}_{i,z})\tau_i
+\overline{v}_{i,z}(T_{\rm p}-\tau_i)\,,
\label{eq:JiDynamic}
\end{align}
considering only targets reachable within the horizon, $\tau_i\le T_{\rm p}$. 
The predicted displacement for remaining with the baseline strategy, $J_{\rm stay}$ in Eq.~(\ref{eq:Jstay}), is unchanged.

\begin{figure}
\includegraphics[width=\linewidth]{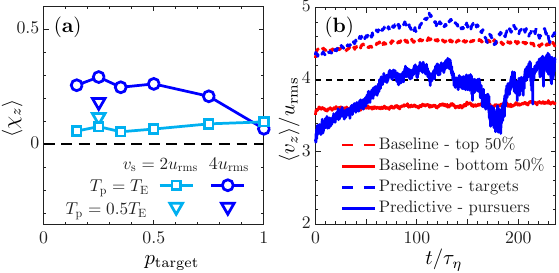}
\caption{
\label{fig:vzMeanDynamic}
Swarm navigation of $1000$ swimmers, with a fraction $p_{\rm target}$ of targets and $1-p_{\rm target}$ of pursuers.
({\bf a}) Averaged vertical velocity enhancement $\langle\chi_z\rangle$ against $p_{\rm target}$ for $\vs=2u_{\rm rms}$ ($\circ$, light blue) and $4u_{\rm rms}$ ({\tiny$\square$},blue) with $T_{\rm p}=T_{\rm E}$, and for $T_{\rm p}=0.5T_{\rm E}$ ($\triangledown$). 
({\bf b}) Evolution of $\langle v_z(t)\rangle$ for $\vs=4u_{\rm rms}$ and $p_{\rm target}=0.5$, averaged over the targets (dashed blue) and pursuers (blue).
For comparison, $1000$ baseline swimmers are split into the instantaneous top $50\%$ (dashed red) and bottom $50\%$ (red) ranked by $u_z$.
}
\end{figure}

Figure~\ref{fig:vzMeanDynamic}({\bf a}) shows $\langle\chi_z\rangle$ against $p_{\rm target}$ with $T_{\rm p}=T_{\rm E}$. As $p_{\rm target}\to 1$, all swimmers become targets, following the baseline, so $\langle\chi_z\rangle$ approaches the baseline value. For $\vs=4u_{\rm rms}$, swarm navigation substantially outperforms the baseline over a broad range of $p_{\rm target}$, reaching about $75\%$ of the enhancement obtained using precalculated targets [Fig.~\ref{fig:vzMeanGlobal}({\bf a})]. 
For $\vs=2u_{\rm rms}$, however, it remains slightly below the baseline.
Unlike the local strategies, the guided strategies were not systematically optimized. 
To illustrate the resulting parameter sensitivity, Fig.~\ref{fig:vzMeanDynamic}({\bf a}) also shows results for $T_{\rm p}=0.5T_{\rm E}$. Thus, the results in Figs.~\ref{fig:vzMeanGlobal}({\bf a}) and \ref{fig:vzMeanDynamic}({\bf a}) could be improved by tuning model parameters, refining the prediction scheme, or selecting better target trajectories.

Figure~\ref{fig:vzMeanDynamic}({\bf b}) shows the evolution of $\langle v_z\rangle$ for targets and pursuers at $\vs=4u_{\rm rms}$ and $p_{\rm target}=0.5$. Both increase during an initial transient and exceed their corresponding baseline populations, obtained by splitting baseline swimmers into the instantaneous top and bottom halves ranked by $u_z$. 
Remarkably, even targets benefit despite following baseline control. Dynamic exchange between target and pursuer roles therefore enhances vertical transport of the entire swarm.

These results demonstrate that precalculated trajectories are unnecessary when swimmers can reach dynamically selected targets sufficiently rapidly. 
At large $\vs$, swarm navigation keeps much of the benefit of precalculated trajectories using only the instantaneous swarm state, while the lack of improvement at $\vs=2u_{\rm rms}$ highlights the importance of interception time.

\section{Conclusions}
We investigated vertical navigation strategies in multiscale turbulence using local flow information, precalculated high-performing trajectories, and dynamic guidance within a swarm.

Local strategies give velocity enhancements of about $10\%$ of $u_{\rm rms}$ over a range of swimming speeds. 
In particular, heuristic clamped gradient ascent performs nearly as well as the previously proposed surfing strategy while using a simpler local signal, showing that different local strategies can exploit the flow with comparable efficiency.

Guidance by precalculated high-performing trajectories yields substantially larger enhancements, reaching about $40\%$ of $u_{\rm rms}$. At small swimming speeds, however, interception becomes slow and can reduce performance below the best local strategy.

Finally, precalculated trajectories are not essential: high-performing swimmers can be identified dynamically and pursued within a swarm. At sufficiently large swimming speeds, this retains a substantial fraction of the benefit of precalculated targets. 
Overall, local flow information provides moderate navigation gains, whereas information about favorable trajectories enables substantially larger gains. Swarms offer a way to acquire and exploit this information dynamically, without prior knowledge of favorable trajectories.

\section{Discussion}
Performance of guided navigation depends on both the quality and accessibility of the target population. Here, targets were selected from baseline trajectories whose mean enhancement $\chi_z$ exceeded a prescribed threshold. 
This criterion does not account for the finite time required to intercept a target. For a given $\vs$, target selection could instead be based on the typical interception time $T_{\rm int}$, favoring trajectories that remain high-performing when they are likely to be closely tracked.

Target accessibility strongly affects $T_{\rm int}$, which decreases with swimming speed and is typically shorter for greedy control [Figs.~\ref{fig:vzMeanGlobal} and \ref{fig:statisticsGlobal}]. 
During interception, swimmers need not sample strong upwelling regions, explaining the reduced benefit of guidance at small $\vs$.
A denser target population would reduce typical interception distances, while greater diversity could increase the chance of finding an accessible high-performing target.
We increased diversity by combining trajectories generated using different baseline strategies and control parameters, but correlations remain. 
Including trajectories with fixed random swimming directions could broaden the target population. The wide distributions in Fig.~\ref{fig:vzMeanLocal}({\bf c},{\bf d}) suggest that such trajectories can occasionally perform well.
Similarly, swarms with more swimmers could provide denser target populations, improving accessibility and reducing interception times.

The benefit of guidance is determined by the ratio of the interception time $T_{\rm int}$ to the observation time $T$. 
For $T_{\rm int}\gtrsim T$, little time remains to exploit a target, whereas for $T_{\rm int}\ll T$, guided performance approaches that of the target. 
Increasing $T$ allows more tracking, but also narrows the finite-time distribution of $\chi_z$, making persistently high-performing trajectories rarer. 
Target selection should therefore consider performance only over the exploitable part of a trajectory. 
By contrast, when $T_{\rm int}\ll T$, trajectories could be divided into shorter segments and $\chi_z$ evaluated for each. Shorter intervals yield larger fluctuations in $\chi_z$, potentially providing more high-performing targets.
This introduces an interception-exploitation tradeoff: frequent target switching provides access to transiently favorable trajectories but increases the fraction of time spent intercepting rather than tracking them.

Our model makes several simplifying assumptions. 
We neglect translational inertia and assume instantaneous reorientation, appropriate for small neutrally buoyant swimmers with reorientation times much shorter than the Kolmogorov time $\tauK$~\cite{voth2017anisotropic,qiu2022active}.
We also neglect hydrodynamic interactions and swimmer feedback on the flow, as appropriate for small, dilute swimmers~\cite{saintillan2013active}. Within the tracking distance $10\etaK$, swimmers pursue targets greedily. This approach should work best for $\vs\gg u_\eta$ (our smallest speed is $\vs=4u_\eta=0.22u_{\rm rms}$). At smaller $\vs$, more sophisticated short-range tracking may be required~\cite{calascibetta2023optimal,koh2025physicsguided}.

Several questions remain open. 
It is unclear whether the advantage of trajectory guidance relies on multiscale turbulence or persists in single-scale flows, and spatial inhomogeneity or coherent structures may change the relative value of local and nonlocal information.
Adaptive target selection, repeated switching, and improved tracking offer natural routes to better performance. Learning-based strategies could further decide when to intercept, abandon, or switch targets based on the instantaneous swimmer and target states. 
More broadly, our results suggest that turbulent navigation depends not only on swimming and local sensing, but also on access to information about favorable trajectories elsewhere in the flow. Understanding how such information can be acquired, communicated, and exploited is a natural step toward collective navigation in turbulence.

\acknowledgments
We acknowledge financial support from Vetenskapsrådet, Grants No. 2018-03974, No. 2023-03617, and from the Knut and Alice Wallenberg Foundation, Grant No. 2019.0079.
Natural language processing (ChatGPT Edu, OpenAI) was used to improve spelling, grammar, punctuation, and textual flow. It was not used to generate or modify the scientific content.

\makeatletter
\def\jan{} \def\feb{} \def\mar{} \def\apr{}
\def\may{} \def\jun{} \def\jul{} \def\aug{}
\def\sep{} \def\oct{} \def\nov{} \def\dec{}
\makeatother

\bibliographystyle{abbrv}
\bibliography{biblio}

\end{document}